\documentclass[11pt]{article}

\usepackage{multirow}
\usepackage{graphics}
\usepackage{subfigure,rotating}
\usepackage{mathtools}
\usepackage{amstext,amsmath,amssymb}
\usepackage{gensymb}
\usepackage{ifsym}
\usepackage[affil-it]{authblk}

\usepackage[left=0.9in,top=0.9in,bottom=1.0in,right=0.9in]{geometry}
\graphicspath{{./figures/}}

\title{\large \bf Neutrino mass hierarchy determination and other physics
  potential of medium-baseline reactor neutrino oscillation
  experiments \footnote{Submitted to the Snowmass 2013 Proceedings}}
\author[13]{\fontsize{10}{10}\selectfont A.B.~Balantekin}
\author[14,13]{H.~Band}
\author[7]{R.~Betts}
\author[13]{J.J.~Cherwinka}
\author[11]{J.A.~Detwiler}
\author[5]{S.~Dye}
\author[14,13]{K.M.~Heeger}
\author[3]{R.~Johnson}
\author[1]{S.H.~Kettell}
\author[6]{K.~Lau}
\author[4]{J.G.~Learned}
\author[2]{C.J.~Lin}
\author[1]{J.J.~Ling}
\author[3]{B.~Littlejohn}
\author[6]{D.W.~Liu}
\author[2]{K.B.~Luk}
\author[4]{J.~Maricic}
\author[9]{K.~McDonald}
\author[12]{R.D.~McKeown}
\author[10]{J.~Napolitano}
\author[8]{J.C.~Peng}
\author[1]{X.~Qian}
\author[11]{N.~Tolich}
\author[12]{W.~Wang}
\author[7]{C.~White}
\author[1]{M.~Yeh}
\author[1]{C.~Zhang}
\author[11]{T.~Zhao}
\affil[1]{\fontsize{10}{10}\selectfont Brookhaven~National~Laboratory, Upton, NY, USA}
\affil[2]{University~of~California and Lawrence~Berkeley~National~Laboratory, Berkeley, CA, USA}
\affil[3]{University~of~Cincinnati, Cincinnati, OH, USA}
\affil[4]{University of Hawaii, Honolulu, HA, USA}
\affil[5]{Hawaii Pacific University, Kaneohe, HA, USA}
\affil[6]{University~of~Houston, Houston, TX, USA}
\affil[7]{Illinois~Institute~of~Technology, Chicago, IL, USA}
\affil[8]{University~of~Illinois~at~Urbana-Champaign, Urbana, IL, USA}
\affil[9]{Princeton~University, Princeton, NJ, USA}
\affil[10]{Rensselaer~Polytechnic~Institute, Troy, NY, USA}
\affil[11]{University of Washington, Seattle, WA, USA}
\affil[12]{College~of~William~and~Mary, Williamsburg, VA, USA}
\affil[13]{University~of~Wisconsin, Madison, WI, USA}
\affil[14]{Yale~University, New Haven, CT, USA}

\date{}
\begin{document}

\maketitle

\begin{abstract}
Medium-baseline reactor neutrino oscillation~(MBRO) experiments
have been proposed to determine the neutrino mass hierarchy (MH) 
and to make precise measurements of the neutrino oscillation 
parameters. With sufficient statistics, better than $\sim 3\%
/ \sqrt{E(MeV)}$ energy resolution and well understood energy non-linearity,
MH can be determined by analyzing oscillation signals driven by the
atmospheric mass-squared difference in the survival spectrum of reactor 
antineutrinos. With such high performance MBRO detectors,
oscillation parameters, such as $\sin^22\theta_{12}$, $\Delta m^2_{21}$, and 
$\Delta m^2_{32}$, can be measured to sub-percent level, which 
enables a future direct unitarity test of the PMNS matrix to $\sim$1\% level
and helps the forthcoming neutrinoless double beta decay
experiments to constrain the allowed $\langle m_{\beta \beta} \rangle$ values. 
Combined with results from the next generation long-baseline beam
neutrino and atmospheric neutrino oscillation experiments, the MH
determination sensitivity can reach 
higher levels. In addition to the neutrino oscillation physics, MBRO
detectors can also be utilized to study geoneutrinos, astrophysical
neutrinos and proton decay. We propose to start a U.S. R\&D
program to identify, quantify and fulfill the key challenges essential
for the success of MBRO experiments.
\end{abstract}

\newpage
\tableofcontents

\section{Introduction}

The precise measurement of $\sin^{2}2\theta_{13}$ by the current generation
of short-baseline reactor neutrino 
experiments~\cite{An:2012eh,An:2012bu,Ahn:2012nd} 
has provided a unique opportunity  
to determine the neutrino mass hierarchy~(MH) in a medium-baseline reactor
neutrino oscillation~(MBRO)
experiments~\cite{Petcov:2001sy,Choubey:2003qx,deGouvea:2005hk,Learned:2006wy,
  Minakata:2007tn, Parke:2008cz, Zhan:2008id, Zhan:2009rs, Ciuffoli:2012iz,
  Ge:2012wj, Capozzi:2013psa}. By
employing large liquid-scintillator~(LS) detectors at distances
greater than $\sim$30 km from nuclear reactors, we can observe the
oscillation signals driven by both 
the solar mass-squared splitting ($\Delta m_{21}^{2}$) and the atmospheric
mass-squared splitting ($\Delta m_{32}^{2}$) in the antineutrino 
energy spectrum~\cite{Zhan:2009rs}. 
The oscillation resulted from the atmospheric mass-squared splitting manifests
itself in the energy spectrum as multiple cycles which 
shift in the opposite directions for inverted hierarchy~(IH) and normal
hierarchy~(NH), as shown in the following formula, 
\begin{eqnarray}
P_{\bar{\nu}_{e}\rightarrow\bar{\nu}_{e}} & = &
1-\sin^{2}2\theta_{13}(\cos^{2}\theta_{12}\sin^{2}\Delta_{31}+\sin^{2}\theta_{12}\sin^{2}{\Delta_{32}})-\cos^{4}\theta_{13}\sin^{2}2\theta_{12}\sin^{2}\Delta_{21}\label{eq:osc-mh}\\
 & = & 1-2s_{13}^{2}c_{13}^{2}-4c_{13}^{2}s_{12}^{2}c_{12}^{2}\sin^{2}\Delta_{21}+2s_{13}^{2}c_{13}^{2}\sqrt{1-4s_{12}^{2}c_{12}^{2}\sin^{2}\Delta_{21}}\cos(2\Delta_{32}\pm\phi),\nonumber 
\end{eqnarray}
where $\Delta_{21}\equiv\Delta m_{21}^{2}L/4E$, $\Delta_{32}\equiv\Delta
m_{32}^{2}L/4E$, 
in which $L$ is the baseline and $E$ is the antineutrino energy, and 
\[
\sin\phi=\frac{c_{12}^{2}\sin2\Delta_{21}}{\sqrt{1-4s_{12}^{2}c_{12}^{2}\sin^{2}\Delta_{21}}}\,,\,\cos\phi=\frac{c_{12}^{2}\cos2\Delta_{21}+s_{12}^{2}}{\sqrt{1-4s_{12}^{2}c_{12}^{2}\sin^{2}\Delta_{21}}}.
\]
In contrast to electron-neutrino appearance experiments 
such as the Long Baseline Neutrino Experiment (LBNE) and the Long Baseline
Neutrino Oscillation Experiment
(LBNO)~\cite{Akiri:2011dv,Bertolucci:2012fb,Diwan:2013eha}, 
which have to take into account the effects from $\delta_{CP}$, MBRO
experiments are free of any effects due to the unknown $\delta_{CP}$
phase. The amount of the  shift in the neutrino energy spectrum due to
different MH is characterized by  the  
ratio of $\Delta m_{21}^{2}/\Delta m_{32}^{2}$, which is about 0.03, 
therefore to make a meaningful measurement of the neutrino MH effect in 
MBRO experiments, one needs excellent energy resolution and well-calibrated
detector energy response. Recent studies show that with a detector energy
resolution $\sim 3\%/\sqrt{E(MeV)}$\footnote{Besides the dominant term due to
photo-electron statistical fluctuation, energy resolution expression
also includes non-uniformity and noise contributions. The overall
resolution needs to be better than 3\% at 1MeV.} and energy non-linearity 
measured to sub-1\% over the entire reactor antineutrino energy
spectrum, a $\Delta\chi^{2}=16$\footnote{See
  Sec.~\ref{sub:Sensitivity-to-mh} on the relationship between $\Delta
  \chi^2$ values and confidence levels for MH determination.}
measurement can be made in 5 years with an exposure of 800
kt-GW$_{th}$ per year at a baseline of
$\sim$60km~\cite{Li:2013zyd}. Such high performance detectors can also
make precise measurements of $\sin^{2}2\theta_{12}$, $\Delta
m_{21}^{2}$and $\Delta m_{32}^{2}$ to sub-percent level. In addition,
they are excellent detectors for studying other important physics
topics such as geoneutrinos, solar neutrinos, atmospheric neutrinos,
and proton decay. Together with the improving $\sin^22\theta_{13}$
precision by the current generation short-baseline reactor neutrino
experiments, it will enable a future direct  unitarity test of the PMNS
matrix to $\sim$1\% level. It can also help to constrain the allowed region
in the phase space of $\langle m_{\beta \beta} \rangle$ vs. $m_{light}$,
the lightest neutrino mass, and provide more precise absolute neutrino
mass constraints should the neutrinoless double beta decay experiments
observe any signals~\cite{Rodejohann:2012xd}. If we are lucky enough
to witness a supernova within 10 $kpc$ during the experiment's live
time, we would expect to record about 6000 supernova neutrinos with
accurately measured energy and time profile for a 20kt detector.

A MBRO experiment at Jiangmen Underground Neutrino Observatory~(JUNO) in
China has been proposed to measure the neutrino mass hierarchy 
~\cite{Li:2013zyd}. The proposed detector contains about 20kt of
liquid scintillator (LS) under a $\sim$700m overburden and will be
located $\sim$60km from two nuclear power plants with a total power of
$\sim$36GW$_{th}$ currently under construction. This experiment is expected
to record $\sim10^{5}$ inverse beta decay~(IBD) events in five
years. Another MBRO experiment named RENO-50 has been proposed in South
Korea. RENO-50 proposes to build a 18kt LS detector at a distance of
47km from the current RENO reactor complex~\cite{RENO-50}.

There are many challenges to such an experiment.
In addition to the requirements of a very large detector volume and very good
energy resolution, very precise energy calibration is also critical for
the experiment. Excellent energy resolution can be achieved with maximal
photocathode coverage in the detector, enhanced scintillation light yield
as well as long scintillator light attenuation length. Calibration
of such a large detector to the required precision is non-trivial. The
energy scale requirements demand the deployment of a comprehensive suite of
calibration sources and making detector response measurement to sub-1\%
level, which pose great challenges in engineering as well as  detector
simulation. These challenges require an intensive, target oriented R\&D
program. US groups with extensive experiences in solar, reactor and
atmospheric neutrino experiments such as Super-K, SNO, KamLAND and Daya Bay
are in an excellent position to undertake such a R\&D program for MBRO
experiments. In this paper, we explore the rich physics potential of
MBRO experiments based on the realistic performance
obtained or extrapolated from the past experiments. By
quantifying the requirements, we are able to identify the key issues that 
the U.S. R\&D program needs to address. The paper is organized as follows.
In Sec.~\ref{sec:Medium-baseline-reactor-neutrino}, we discuss the
nature of the MH signals in massive LS detectors and evaluate the
sensitivities based on the current proposals and inputs from the existing
reactor neutrino experiments.
Precision oscillation parameter measurements are discussed in
Sec.~\ref{sec:Precision-measurement}. Based on our sensitivity studies and
proposed physics goals, we identify the key detector performance
requirements and discuss their realization and possible R\&D plans in
Sec.~\ref{sec:Performance-requirements}. Physics potential other than
neutrino oscillation physics is covered in
Sec.~\ref{sec:Other-potential-physics}. We then give our summary and
conclusions on MBRO experiments and possible U.S. roles in such
experiments.

\section{\label{sec:Medium-baseline-reactor-neutrino}Medium-baseline reactor
neutrino experiments resolving MH}

\subsection{MH signal in MBRO experiments and challenges}

As has been pointed out in the previous section, it is plausible to determine
the MH from the energy spectrum of detected reactor antineutrinos at
suitable distances from a nuclear
reactor~\cite{Petcov:2001sy,Choubey:2003qx,deGouvea:2005hk}. As shown in
Eq.~\ref{eq:osc-mh}, the MH dependence comes solely through the phase shift
$\phi$, which takes a plus sign for NH and minus sign for IH. The value of
$\phi$ depends on the neutrino energy  and is in general small ($\sim$5\% of
$\Delta m^2_{32}$) in the energy  range of the reactor neutrinos (1.8 - 10
MeV). To resolve this  small spectrum difference between the two MH
hypotheses, it requires both large statistics and good control of
systematics.

\begin{figure}
\begin{centering}
\includegraphics[height=2in]{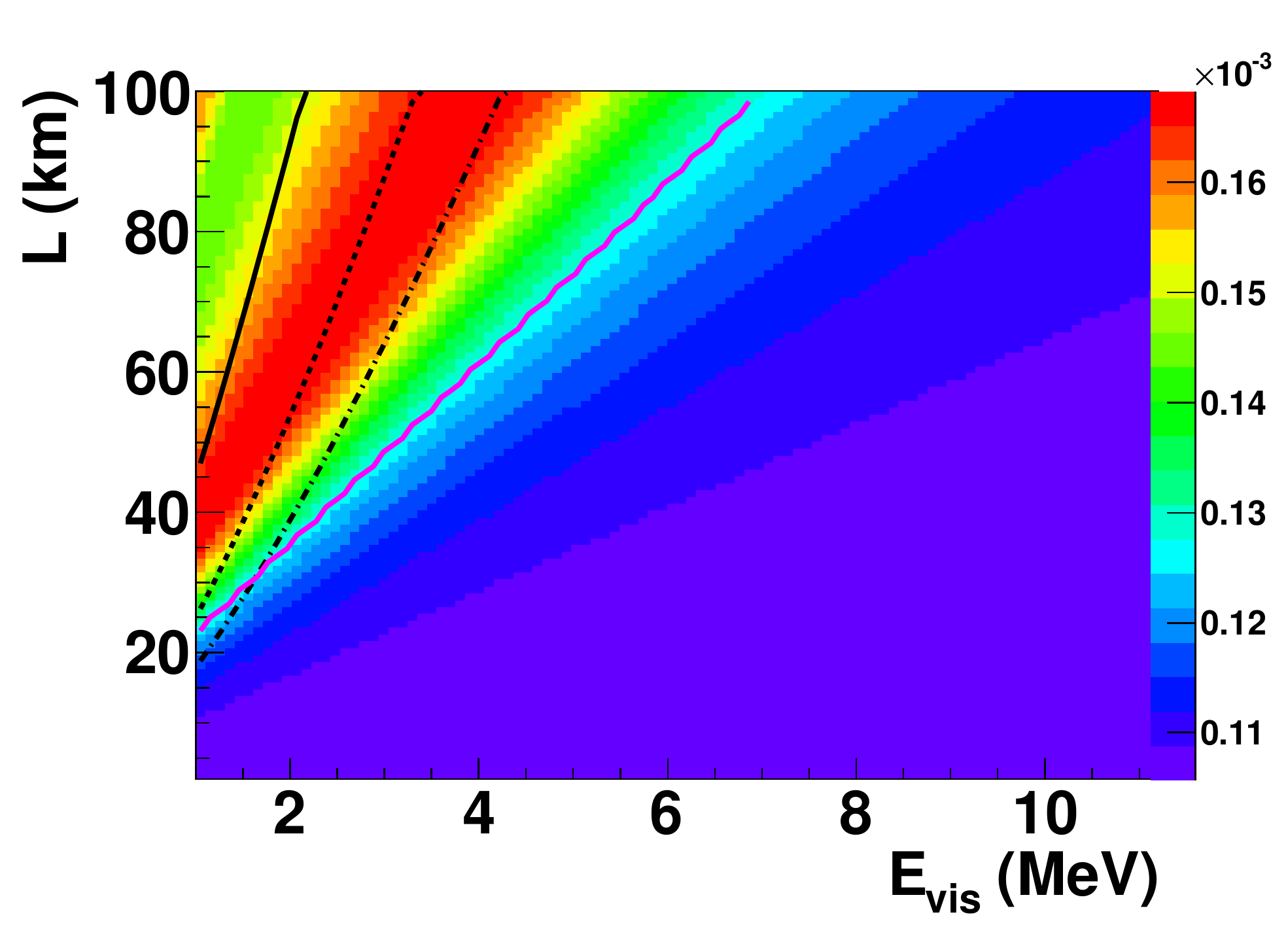}
\includegraphics[height=1.8in]{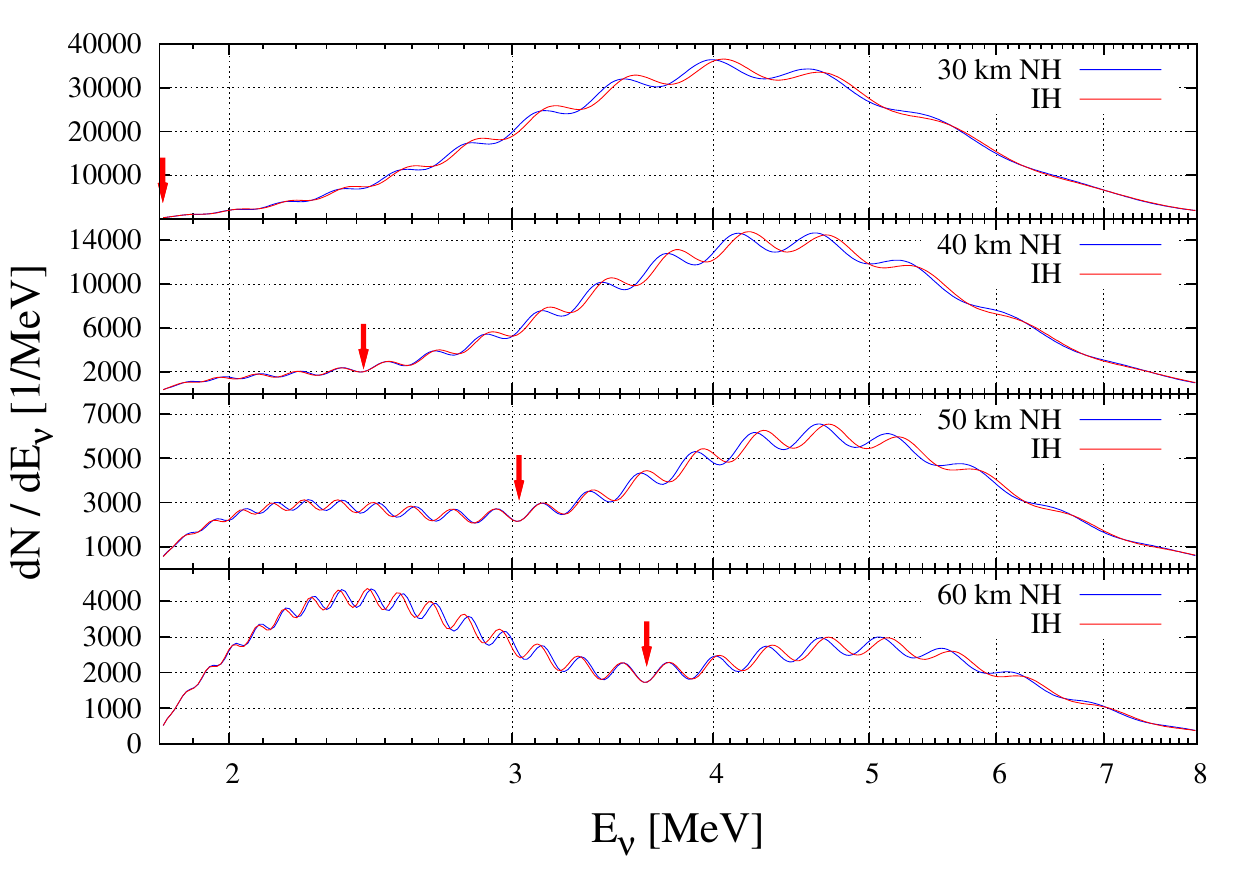}
\par\end{centering}
\caption{\label{fig:The-effiective-shift}Left: the effective mass-squared
difference shift $\Delta m^{2}_{\phi}$ as a function of baseline and visible
prompt energy $E_{vis} \approx E_{\nu} - 0.8~MeV$.  Right: the comparison of
energy spectra between NH and IH at different baselines. Arrows mark the
location where $\Delta m_{\phi}^2 = 0$.}
\end{figure}

Ref.~\cite{Learned:2006wy,Parke:2008cz,Zhan:2008id,Zhan:2009rs,Qian:2012xh}
show that energy resolution better than $\sim 3\% / \sqrt{E(MeV)}$ is
needed in order to resolve the difference between NH and IH. This
can  be easily understood from the left panel of
Fig.~\ref{fig:The-effiective-shift}, which shows the energy and
baseline dependent phase shift of $\phi$. In principle, the MH can be
resolved by comparing the measured effective mass-squared difference $\Delta
m^2_{32} \pm \Delta m^2_{\phi}/2$ at low energy ($\sim$ 3 MeV) vs. 
that at high energy ($\sim$ 6 MeV). Here, we have defined $\Delta m^2_{\phi}
\equiv 4E\phi/L$ based on Eq.~\ref{eq:osc-mh}. For NH, the effective
mass-squared difference at low energy will be larger than that at high
energy and vice versa for IH. However, at low energy, since the $L/E$ is
large, a poor energy resolution will smear the oscillation signals
corresponding to the atmospheric oscillation ($\Delta m^2_{32}$), leading to
difficulties for measuring the true oscillation frequency at low energy
region.

\begin{figure}[htp]
\centering{}\includegraphics[width=0.9\textwidth]{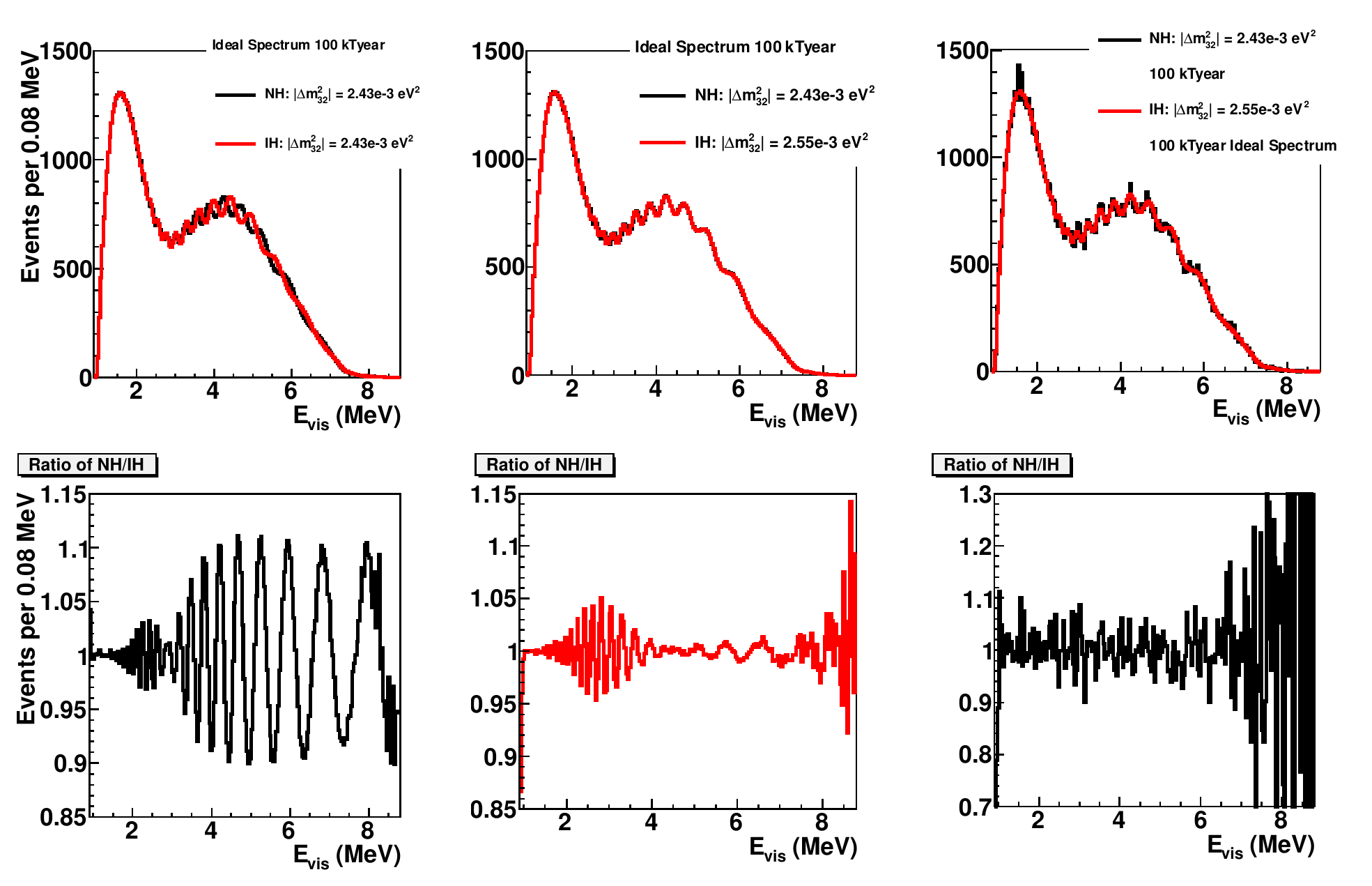}
\caption{\label{fig:nh-ih-ratio}Comparison of the energy spectra of the
  positron signal under different assumptions for NH and IH
  cases. The uncertainty in $\Delta m_{32}^{2}$ was assumed as $\sim
    0.13\times10^{-3}eV^{2}$ in Summer 2012 when the paper was
    published.}
\end{figure}

Besides the challenges held in the energy resolution, 
uncertainties in the value of $\Delta m_{32}^{2}$ could result in the
degeneracy of NH and IH that makes it impossible to measure MH.
Fig.~\ref{fig:nh-ih-ratio} shows comparison of the visible energy
spectra of the inverse beta decay~(IBD) events between NH and IH at a
distance of 60km. The left panel is the ideal case and the difference
between NH and IH is visible across the entire spectrum.
Due to the $\Delta m_{32}^{2}$ uncertainty, as shown in the middle panel of
Fig.~\ref{fig:nh-ih-ratio}, degenerated oscillation probabilities
significantly reduce the spectrum difference between different MHs. The
situation becomes worse when statistical fluctuations are included, as shown
in the right panel of Fig.~\ref{fig:nh-ih-ratio}.
To overcome this challenge, we need more accurate independent
measurement of $\Delta m_{32}^{2}$ as pointed out in
Ref.~\cite{Li:2013zyd, Qian:2012xh}. Reference~\cite{Li:2013zyd} shows that,
under different set of assumptions, with the possible improvement on
$\Delta m_{32}^{2}$ to 1.5\% (T2K, NO$\nu$A), the MH sensitivity can be
increased to $\Delta\chi^{2}\cong20$ in 6 year running time.

The baseline of MBRO experiments needs to be greater than $\sim$30km 
in order to resolve the MH signal reliably. This is illustrated in
Fig.~\ref{fig:The-effiective-shift}, taken from Ref.~\cite{Qian:2012xh}
(left) and Ref.~\cite{Ge:2012wj} (right). From the left panel of
Fig.~\ref{fig:The-effiective-shift}, for baselines less than
$\sim$30km, $\Delta m_{\phi}^{2}$ is rather uniform across the
entire IBD spectrum and the current uncertainty in $\Delta m_{32}^{2}$ 
can easily absorbs it thus it is impossible to distinguish between NH and
IH. For baselines greater than $\sim$30km and close to
$\sim$60km, the solar oscillation suppression of the reactor flux
is near its maximum and energy dependent $\Delta m^{2}_{\phi}$ makes
the MH effect more visible. The right panel of
Fig.~\ref{fig:The-effiective-shift} shows the same observation from the
spectrum perspective, the opposite phase shifts between the low and high
energy regions only appear when baselines are greater than
$\sim$30km.

The needs of good energy resolution and large number of free proton targets
for IBD antineutrino reaction make liquid scintillator~(LS) the
best choice for MBRO experiments. However, LS has a notorious property:
non-linear energy response caused by energy quenching and Cherenkov radiation. 
Combined with possible electronic non-linear effect,
inaccurate energy calibration could potentially cause degenerated
energy spectra between different MHs if the energy reconstruction is biased in
the following non-linear fashion,
\begin{equation}
E_{rec}=\frac{2|\Delta^{\prime}m_{32}^{2}|+\Delta m_{\phi}^{2}(E_{\bar{\nu}_{e}},\, L)}{2|\Delta m_{32}^{2}|-\Delta m_{\phi}^{2}(E_{\bar{\nu}_{e}},\, L)}E_{real}.\label{eq:non-linear-deg}
\end{equation}
Here $E_{rec}$ is the reconstructed energy and $E_{real}$
is the true energy. $|\Delta^{\prime}m_{32}^{2}|$ represents a different
$\Delta m_{32}^{2}$ best-fit value obtained from the observed energy spectrum
allowed by its current uncertainty. It has been illustrated in
Ref.~\cite{Qian:2012xh} that with the allowed uncertainty $\delta (\Delta
m_{32}^{2}) = 0.13\times10^{-3}eV^{2}$, to break the degeneracy, energy
non-linearity needs to be understood to the sub percent level. The current
generation of large LS detectors can achieve a precision of $\sim$2\%. This
requirement can be relaxed if the uncertainty in $\Delta m_{32}^{2}$ get
improved.

In addition to the most critical requirements on energy resolution
and energy response, there are other challenges in MBRO experiments, such as
backgrounds, reactor core distributions and event statistics. We will
discuss these factors with the sensitivity study in the following sections.

\subsection{\label{sub:Sensitivity-to-mh}Mass hierarchy sensitivity study}

\subsubsection{The $\chi_{min}^{2}$ comparison method resolving MH}

To study the physics sensitivity of MH determination in MBRO experiments, a
$\chi^{2}$ is constructed using the pull method to do a model comparison
between NH and IH as follows,

\begin{equation}
\chi^{2}=\sum_{i=1}^{N}2 \cdot (N_i^{exp} - N_i^{obs} + N_i^{obs}\cdot 
\log(N_i^{obs}/N_i^{exp}))+\chi_{penalty}^{2},\label{eq:chi-square}
\end{equation}
where $N_{i}^{obs}$ is the number of observed IBD events in energy
bin $i$ given one of MH is true 
and $N_{i}^{exp}$ is the expected number of IBD events in 
bin $i$ assuming either NH or IH. The penalty component $\chi_{penalty}^{2}$ 
includes systematic constraints and any \emph{a priori} knowledge 
on oscillation parameters from other experiments. 
The best-fit minimal $\chi^{2}$ difference between the
two MH hypotheses is defined as: $\Delta\chi^{2}\equiv\chi_{min,IH}^{2}-\chi_{min,NH}^{2}$.
Naturally, a positive $\Delta\chi^{2}$ indicates the NH model is
preferred by the data over the IH model as the better model has smaller
$\chi_{min}^{2}$. 

\begin{figure}
\begin{centering}
\includegraphics[height=2.5in]{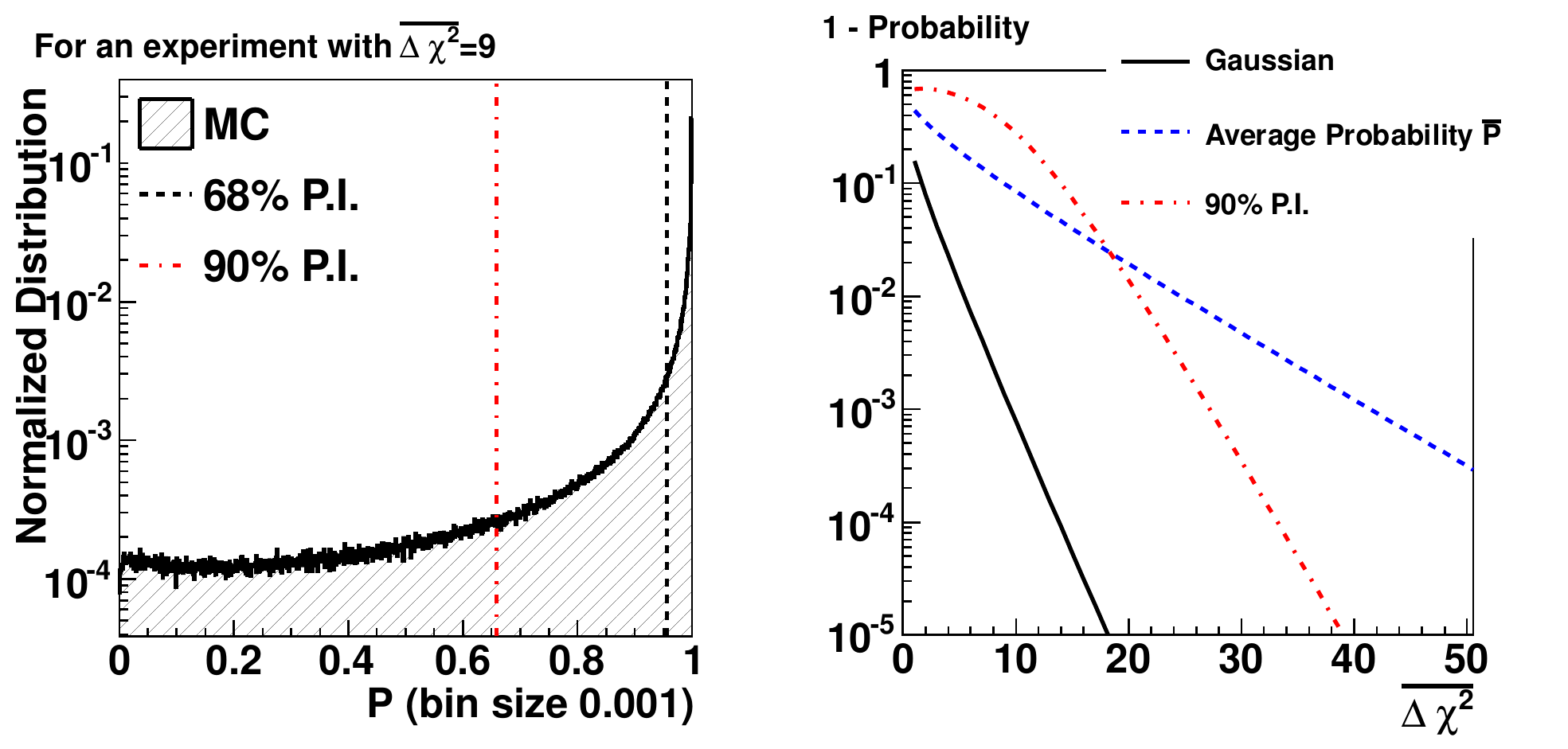}
\par\end{centering}
\caption{\label{fig:cl-vs-deltachi} The left panel shows the distribution of
  $P(NH|x)=P(NH|\Delta\chi^2)$ over the population of potential data
  $x$ that arises from an experiment with $\overline{\Delta \chi^2}=9$
  where the truth is NH. The mean of this distribution is 90.14\%.
  Lower bound of the 68\% and 90\% probability intervals are
  plotted. That is, 68\% (90\%) of the data $x$ would yield a
  $P(NH|x)$ that falls to the right of the dash-dotted (dashed)
  line. These two lines are also commonly referred as the 32th and the
  10th percentile.  The right panel plots several sensitivity metrics
  (subtracted from $1$ for clarity), against $\overline{\Delta
    \chi^2}$ that ranges from 1 to 50. Note that all the lines are
  decreasing because higher values of $\overline{\Delta \chi^2}$
  corresponds to more sensitive experiments.  This is done for three
  different criteria: the Gaussian interpretation (derived from 
  the one-sided p-value with one degree of freedom),
  $\overline{P}$ and $P^{90\%}_{T=NH}$. The Gaussian interpretation is
  seen to be over-optimistic in describing the ability of the
  experiment to differentiate the two hypotheses.}
\end{figure}

For continuous quantities that can be approximated by normal distributions,
the $\sqrt{|\Delta\chi^{2}|}$ in the unit of standard Gaussian
deviation $\sigma$ is commonly used as the confidence level~(C.L.). 
However, as pointed out in Ref.~\cite{Qian:2012zn}, due to the discrete 
nature of MH, the square root rule does not apply any more in setting 
the C.L. for MH measurement. The proper
relation between $\Delta\chi^{2}$ and C.L. for a simple example 
is shown in Fig.~\ref{fig:cl-vs-deltachi}. The new rule requires a greater
$\Delta\chi^{2}$ to reach the same C.L. This fact has
been confirmed in Ref.~\cite{Ge:2012wj, Capozzi:2013psa}. For convenience,
we will still use $\Delta\chi^{2}$ as the quantity representing sensitivity. 
However, we should keep in mind that the C.L. vs $\Delta \chi^2$
relation for MH measurement has been modified. This is a special case of the
Feldman-Cousins method when the distribution of the measured quantity
is better approximated by a 2-value binomial distribution. Based on
this finding, $\Delta\chi^{2}$ has to be $\sim$100 to reach the  5$\sigma$
discovery level, well beyond the sensitivity of any current MBRO experiment
proposals. To reach the 3$\sigma$ strong indication level, a MBRO experiment
needs to be able to reach $\Delta\chi^{2}$ $\sim$36, which is still very
difficult based on current predictions. 

\subsubsection{Background assumptions}

We have taken the background spectra from Daya Bay or theoretical
calculations and extrapolated the background rates based on the KamLAND
results~\cite{An:2012eh,An:2012bu,PhysRevLett.100.221803}. The 
backgrounds we have considered include: 
\begin{itemize}
\item Accidental background \\
  With five years of run time, we expect $\sim$3000 events from the accidental
  background and its prompt signal spectrum is assumed to be the
  same as the one in Daya Bay. Since the prompt signal spectrum
  can be directly measured with high precision, we assume the accidental
  background rate uncertainty is negligible. 
\item  $^9$Li/$^8$He background \\
  We expect $\sim$550 events. A 30\% rate uncertainty in the $^9$Li/$^8$He
  background is assumed. We use the theoretical spectrum in this study.
\item Fast neutron background \\
  We expect $\sim$400 events. The energy spectrum is assumed to
  be flat and a 50\% rate uncertainty is assumed. 
\item $^{13}$C$(\alpha$, n)$^{16}$O background \\
  We expect $\sim$6300 events. The energy spectrum is assumed to be the same
  as that measured in Daya Bay. A 20\% rate uncertainty is assumed.
\item Geoneutrino background \\
  We expect $\sim$3600 events. A 10\% rate uncertainty
  is assumed. We use the theoretical spectrum in this study.
\end{itemize}

For all the backgrounds above, we neglect uncertainties in the spectral shape.

\subsection{Impact of detector energy responses}

\begin{figure}
\begin{centering}
\includegraphics[height=2.5in]{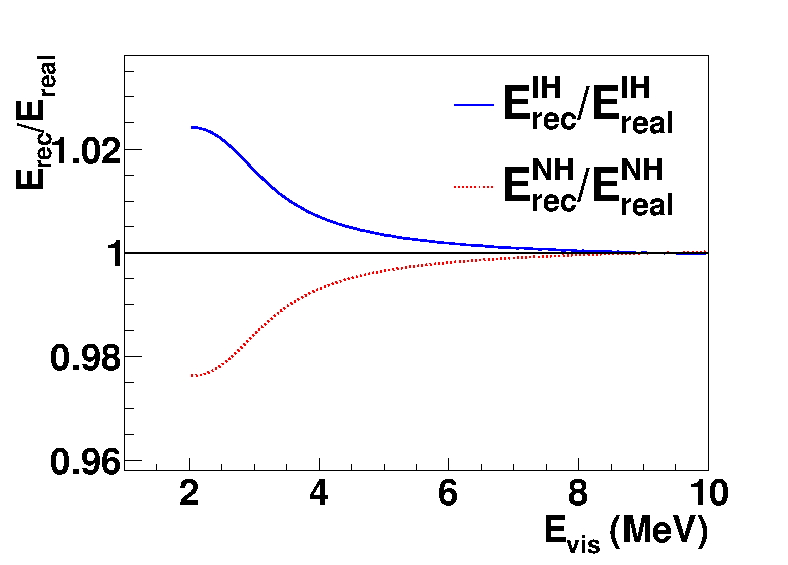}
\par\end{centering}
\caption{\label{fig:non-linear-deg} Energy non-linearity models that can
  cause degeneracy when the true MH is NH (red) and IH (blue). See text for
  more explanations.}
\end{figure}

\begin{figure}
\begin{centering}
\includegraphics[height=2.5in]{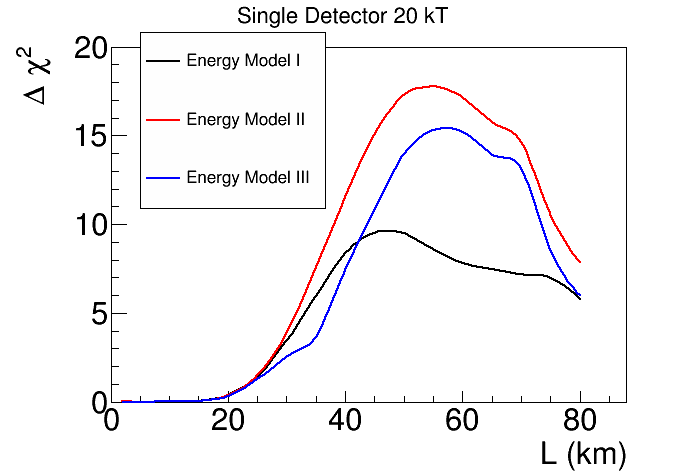}
\par\end{centering}
\caption{\label{fig:sensitivity-baseline}MH sensitivity evolution
  with respect to different baseline choices under different energy
  response assumptions.}
\end{figure}

In order to study the effect of non-linear energy scale uncertainties, we have
assumed 3 types of energy models:
\begin{enumerate}
\item Model I: \\
  The non-linear model set by Eq.~\ref{eq:non-linear-deg}, also shown as the
  red curve in Fig.~\ref{fig:non-linear-deg}. We have assumed NH as the true
  MH and the energy scale uncertainty of 1\%.
\item Model II: \\
  A linear shift in the absolute energy scale uncertainty of 1\%,
  $\sigma_{scale}=1\%$.
\item Model III: \\
  The current preliminary Daya Bay non-linear model, which is a
  combined model based on multiple nearly independent models. These
  multiple energy models are constructed with different emphases of
  the calibration and control data. The energy scale uncertainty is at
  $\sim$1\% level across the IBD spectrum. This approach provides the
  flexibility in the functional form of the non-linearity model. For
  details, see Ref.~\cite{dayabay:shape}.
\end{enumerate}
With the above 3 different energy scale models, we first perform a
baseline scan. Fig.~\ref{fig:sensitivity-baseline}
shows the sensitivity evolution with respect to the baseline. Depending
on the choice of the energy response models, optimal baseline varies
between 40km and 60km, which is consistent with the findings of other groups.

\begin{figure}
\begin{centering}
\includegraphics[height=2.5in]{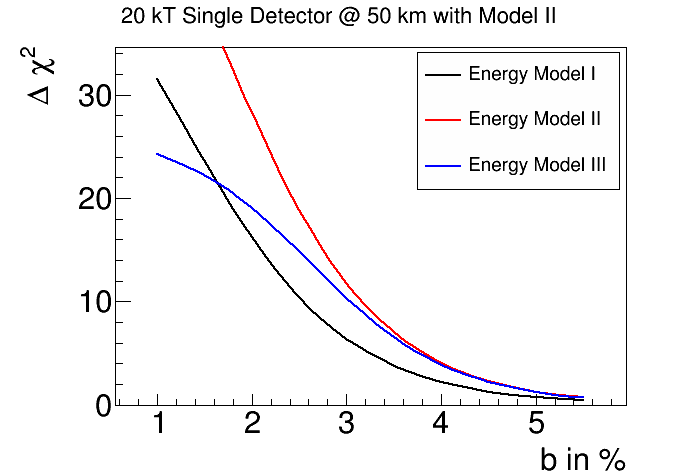}
\par\end{centering}
\caption{\label{fig:sensitivity-res}MH sensitivity as a function
  of the $b$ term in the resolution function (Eq. \ref{eq:deltae_e_def}) for the 3 different energy
  scale models.}
\end{figure}

Next, we study the effect of energy resolution on the sensitivity. For
energy resolution, we choose the following generic model,
\begin{equation}
\frac{\Delta E}{E}=\sqrt{a^{2}+\frac{b^{2}}{E}+\frac{c^{2}}{E^{2}}}.\label{eq:deltae_e_def}
\end{equation}
Where $\Delta E$ is the energy resolution at a total visible energy
$E$, $a$ is due to the energy leakage and detector non-uniformity, 
$b$ is the term that depends on the photo-electron~(PE) statistics, and $c$
is due to background and noise.
We have assumed $a=$0.7\% and $c=0.85$\%, which are extrapolated 
from the performance of the Daya Bay and KamLAND detectors. 
Fig.~\ref{fig:sensitivity-res} shows the sensitivity dependence 
on the statistical uncertainties
in the total number of PEs. As we can see, the sensitivity drops dramatically
once the PE uncertainty is greater than $\sim$3\% for Model
2 and 3. For the designed Model I, the turning point is even lower,
at $\sim$2.5\%.

\subsection{Expected sensitivity of MBRO experiments to MH}

\begin{figure}
\begin{centering}
\includegraphics[height=2.5in]{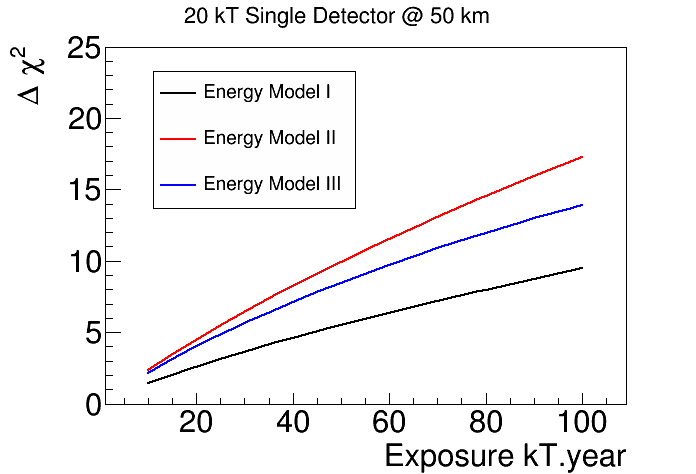}
\par\end{centering}
\caption{\label{fig:The-time-evolution}The time evolution of $\Delta\chi^{2}$
with respect to exposure}
\end{figure}

Fig.~\ref{fig:The-time-evolution} shows the sensitivity evolution
with respect to exposure. We see that with the designed
degeneracy energy scale (energy model I), the $\Delta\chi^{2}$ can reach
$\sim$10 in a 5 year run, which is a very pessimistic
situation. With the current preliminary Daya Bay energy scale model (energy
model III) and uncertainty, the final $\Delta\chi^{2}$ could reach $\sim$14,
which is about 2$\sigma$ ($\Delta \chi^{2} \sim 16$) quoting the
conventional frequentist statement on
the C.L. of the measurement. With a linear energy scale uncertainty
model (energy model II), the 5-year $\Delta\chi^{2}$ could reach
$\sim$17,which corresponds to $\sim$2$\sigma$ C.L
($\Delta \chi^{2} \sim 16$).

\begin{figure}
\begin{centering}
\includegraphics[height=2.5in]{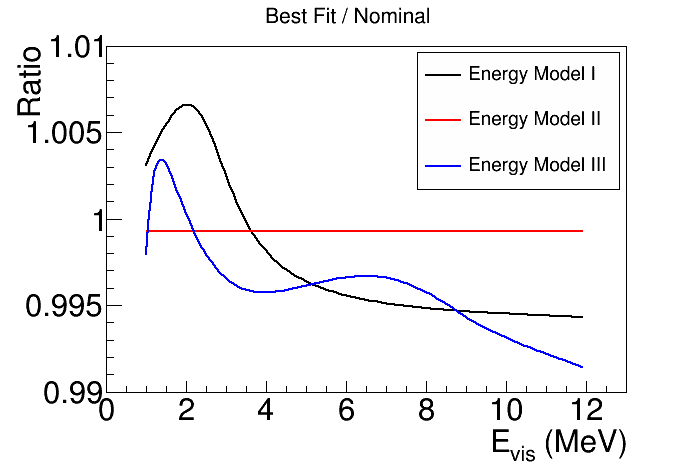}
\par\end{centering}
\caption{\label{fig:bestfit}Fitted energy model at the $\Delta \chi^2$
  minimum for three energy models. }
\end{figure}

As can be seen,  the MH sensitivity strongly depends on the choice of
the non-linearity model (10 vs. 14 vs. 17 for
model I, II, and III,  respectively). Fig.~\ref{fig:bestfit} shows the
fitted energy model at the  $\Delta \chi^2$ minimum for these three
models. Therefore, it is important to have a good understanding of the energy
response, at least the functional form of the energy model. The latter can
also be viewed as a good constraint on the relative energy scale between the
low and high energy regions.

\subsection{Reactor flux uncertainty impact}
The knowledge of the reactor spectrum plays an important role to obtain  
the absolute energy model, which would then help to increase the MH 
sensitivity in MBRO experiments. This is due to that reactor flux is
correlated between energies\cite{Huber:2011wv,Mueller:2011nm}. In this
study, we have used the correlation matrix based on the Daya Bay reactors.
Table~\ref{tab:flux-improvement} shows the improvements on MH sensitivity 
as reactor flux uncertainty is reduced. 
Smaller reactor flux uncertainties can be achieved by employing near
detectors. For example,  RENO-50 is planning to use the current RENO detectors as its near
detectors\cite{RENO-50}. 
\begin{table}
\caption{\label{tab:flux-improvement}Improvements in MH sensitivity with
100kt exposure as a function of improvement in the reactor flux.}
\centering{}%
\begin{tabular}{|c|c|c|c|}
\hline 
Uncertainty improvement & $\Delta\chi^{2}$ (Model I) & $\Delta\chi^{2}$ (Model II) & $\Delta\chi^{2}$ (Model III)\tabularnewline
\hline 
\hline 
Current $\sim$3\% & 9.5 & 17.3 & 13.9\tabularnewline
\hline 
Improve by a factor of 2 & 11.5 & 21.7 & 18.4\tabularnewline
\hline 
Improve by a factor of 3 & 12.1 & 23.2 & 19.9\tabularnewline
\hline 
Improve by a factor of 4 & 12.4 & 23.8 & 20.5\tabularnewline
\hline 
Improve by a factor of 5 & 12.6 & 24.1 & 20.9\tabularnewline
\hline 
\end{tabular}
\end{table}

\subsection{A dual detector design with ratio methods}
With two detectors, one can form ratios between these two detectors, so that 
the uncertainties from the reactor spectrum are largely
canceled\footnote{The assumption here is not to take the theoretical
  uncertainty estimations of reactor spectrum as granted.}. 
However, as shown in Ref.~\cite{Qian:2012xh}, using ratios directly would be 
more sensitive to the uncertainty in the energy model, as the constraint from
the knowledge of the reactor spectrum is not being used. This is also true 
for the proposed Fourier transformation
methods~\cite{Learned:2006wy,Zhan:2008id,Zhan:2009rs}.
On the other hand, Ref.~\cite{Ciuffoli:2012bp}
showed that by placing a second functionally identical detector at
$\sim$30 km baseline, the energy non-linearity requirement can be
greatly relaxed. This is because the MH-dependent oscillation
patterns are different at the two baselines, therefore a single ``wrong''
non-linearity can not fit both detectors if the two detectors have 
highly correlated energy responses. In our sensitivity calculation,
we find that such a configuration of detectors does improve the sensitivity
significantly. Our results are shown in
Table.~\ref{tab:dual-detector-improve}. In our study, we have assumed the
second detector's energy scale is fully correlated with the far
detector. With the assumed energy scale uncertainties based on the current
Daya Bay preliminary results, a second detector at $L$=30km can
significantly improve the MH sensitivity with the ratio method. In order to
reach the same sensitivity of the dual detector configuration, single
detector setup has to be able to reduce the energy scale uncertainties
significantly.
\begin{table}
\caption{\label{tab:dual-detector-improve}Improvement in MH sensitivity
for the degeneracy non-linearity model, Model I, with different second
detector options and under different energy scale uncertainty improvements}
\centering{}%
\begin{tabular}{|c|c|c|}
\hline 
2nd Detector & $\Delta\chi^{2}$ & $\Delta\chi^{2}$
($\sigma_{scale}/4)$\tabularnewline
\hline 
\hline 
20kt at 53km & 4.2 & 14.3\tabularnewline
\hline 
0.1kt at 2km & 4.9 & 11.5\tabularnewline
\hline 
5kt at 30km & 10.3 & 13.6\tabularnewline
\hline 
\end{tabular}
\end{table}

\section{\label{sec:Precision-measurement}Precision measurements and synergy
with $\nu_{\mu}/{\bar{\nu}}_{\mu}$ disappearance experiments}

\subsection{Precision oscillation parameter measurement}

\begin{figure}
\begin{centering}
\includegraphics[width=0.9\textwidth]{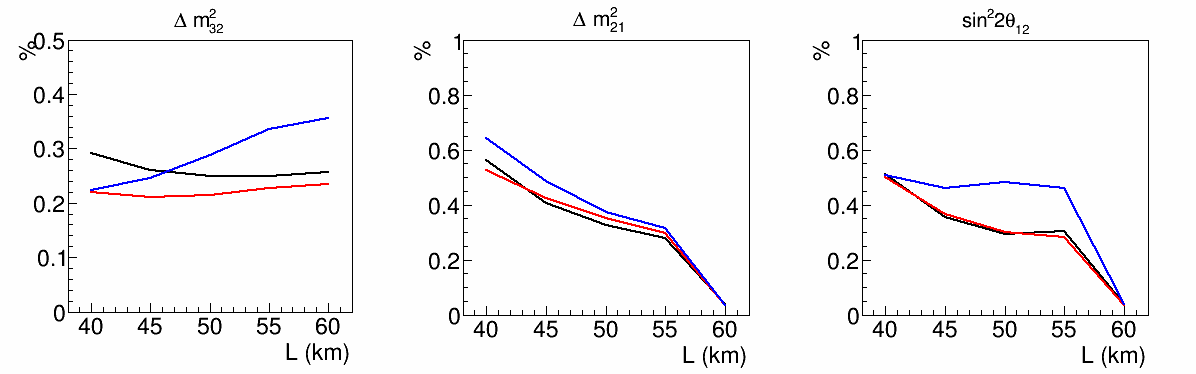}
\par\end{centering}

\caption{\label{fig:precison-baseline}Projected precision oscillation parameter
measurement uncertainties at different baselines. Black: Energy Model
I; Red: Energy Model II; Blue: Energy Model III}
\end{figure}

\begin{figure}
\begin{centering}
\includegraphics[width=0.9\textwidth]{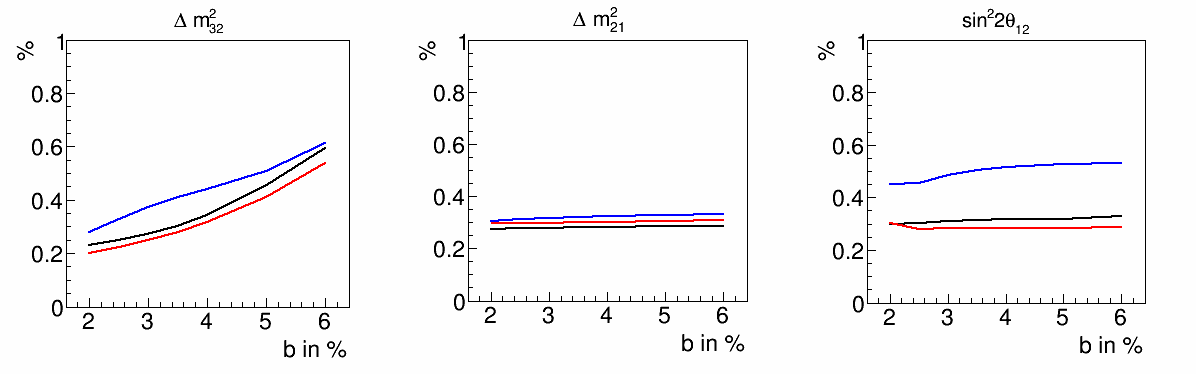}
\par\end{centering}

\caption{\label{fig:precision-resolution}Projected precision oscillation
  parameter measurement uncertainties at different energy resolution
  performances. Black: Energy Model I; Red: Energy Model II; Blue: Energy
  Model III}

\end{figure}

With $\sim$40 detected reactor neutrino events per day, and the multiple
oscillation cycles in the energy range of reactor neutrinos, it is
estimated~\cite{Li:2013zyd} that $\Delta{m}_{21}^{2}$, $\Delta{m}_{31}^{2}$
and $\sin^{2}\theta_{12}$ can be measured to a precision of $\sim$0.6\%.
It enables a future direct unitarity test of the PMNS matrix 
($|U_{e1}|^2+|U_{e2}|^2+|U_{e3}|^2 \stackrel{?}{=} 1$) to a sub-percent
level~\cite{Qian:2013ora}. Our study with background assumptions listed in
Sec.~\ref{sub:Sensitivity-to-mh} shows that sub-percent precision
oscillation parameter measurement is plausible.
Fig.~\ref{fig:precison-baseline} shows the precisions of measuring
oscillation parameters at different baseline values. 
It is not surprising that the best baseline for measuring 
solar sector parameters happens at the solar oscillation maximum, which
is $\sim$60km
for the energy range of reactor flux. On the other hand, the $\Delta
m_{32}^{2}$
precision does not depend on the baseline much as long as there are
multiple oscillation cycles in the survival spectrum. However, the
precision to $\Delta m_{32}^{2}$ decreases quickly as energy resolution
is over $\sim$3\%/$\sqrt{E(MeV)}$ as shown in
Fig.~\ref{fig:precision-resolution}. As the solar scale oscillation spans
over the entire spectrum at the distance of  $\sim$60km, the
energy resolution has little impact to the solar oscillation parameter
precisions.

\subsection{Synergy with $\nu_{\mu}/{\bar{\nu}}_{\mu}$ disappearance
  experiments in MH determination}

\begin{figure}
\begin{centering}
\includegraphics[height=2.5in]{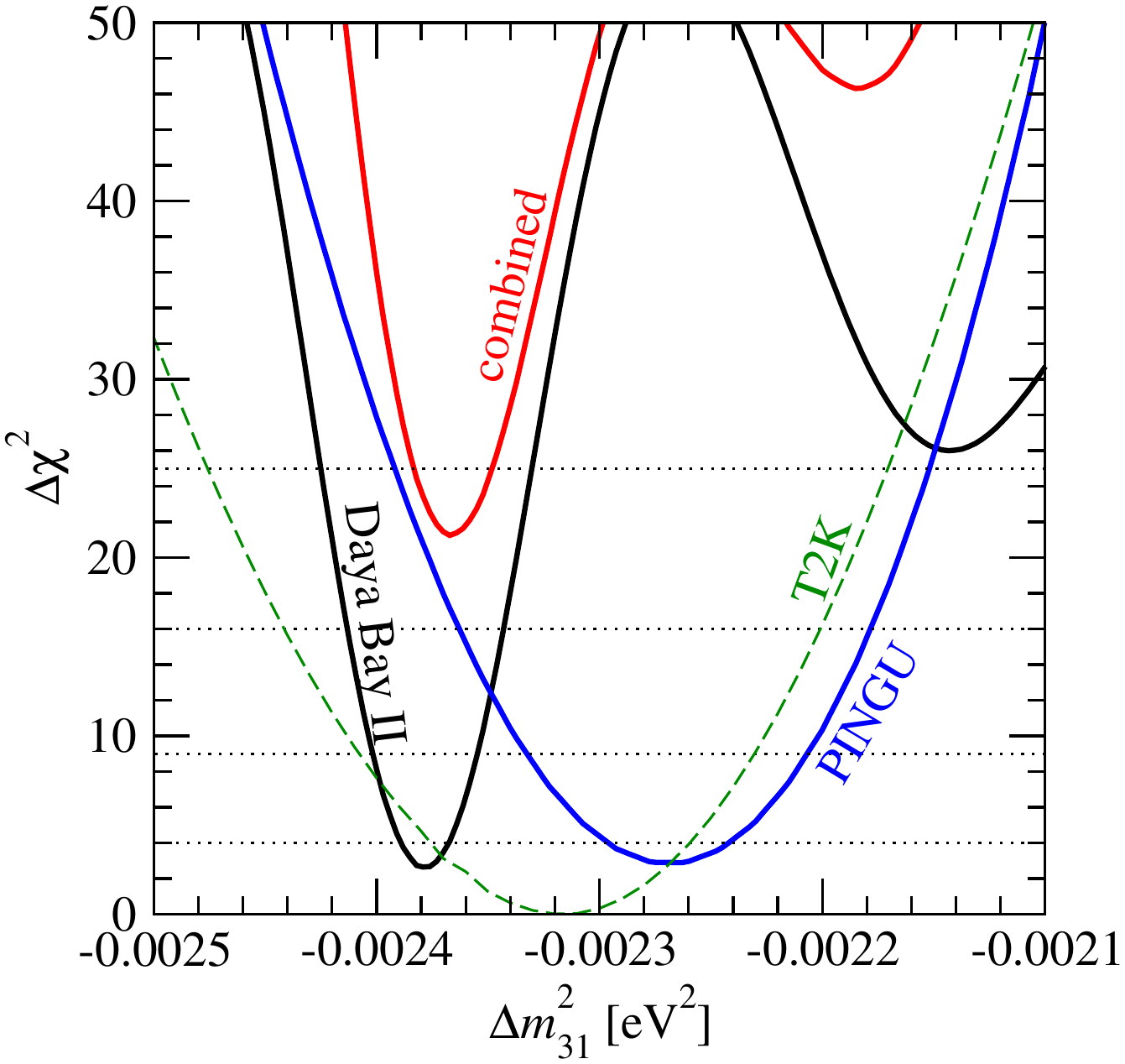}
\includegraphics[height=2.5in]{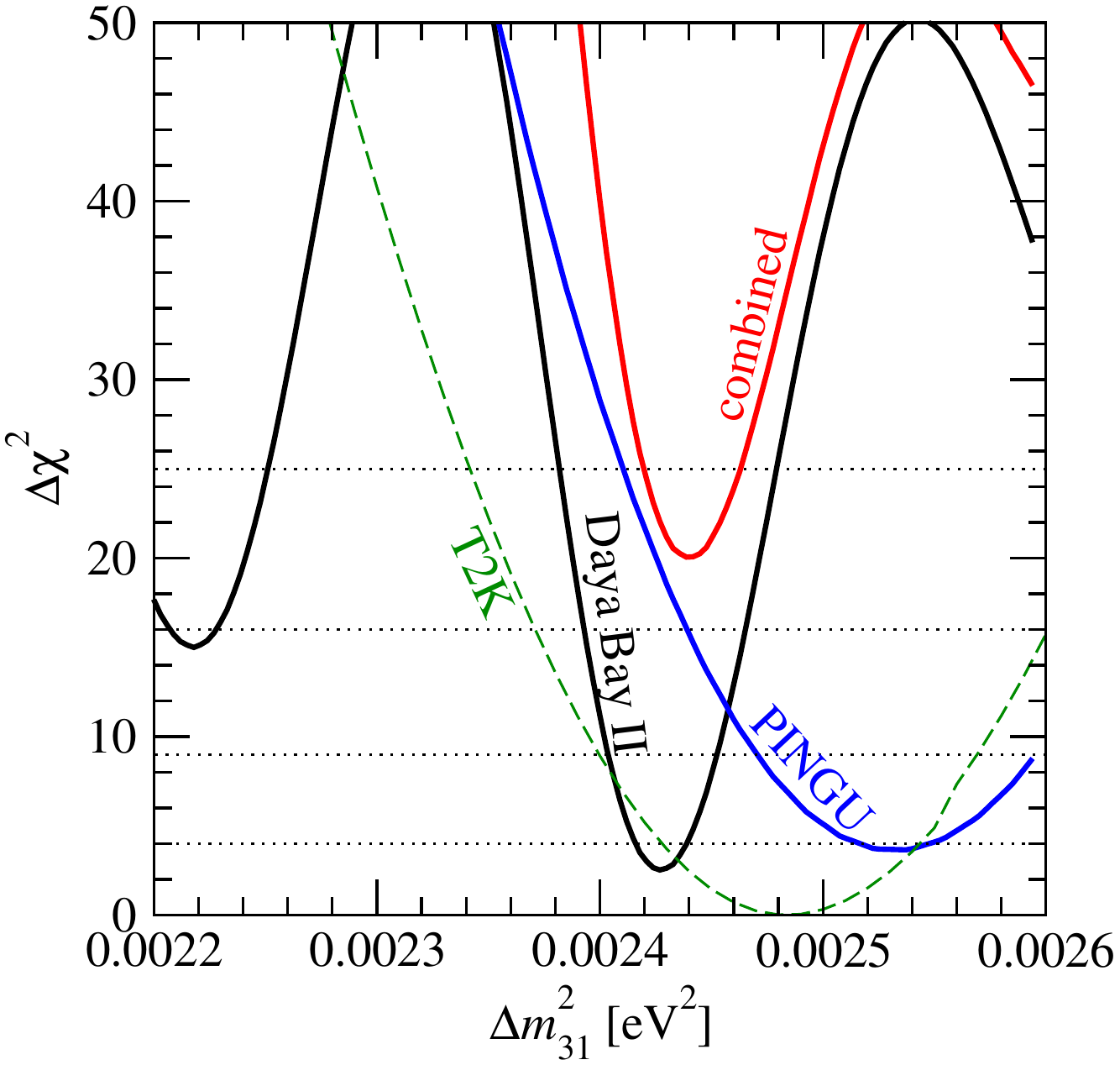}
\par\end{centering}

\caption{\label{fig:Combined-JUNO}Combined JUNO, T2K and PINGU data
  enhancing the sensitivity to the MH. Left panel assumes NH as the true MH
  and the right panel assumes IH.}
\end{figure}

As the uncertainty in $\Delta m^2_{32}$ gets improved in coming years from
$\nu_{\mu}/{\bar{\nu}}_{\mu}$ beam experiments like NO$\nu$A and T2K, the
sensitivity of MBRO experiments also gets improved significantly as shown in
Ref~\cite{Li:2013zyd}.
While MBRO experiments are utilizing vacuum oscillation of reactor
antineutrinos to measure MH, experiments like PINGU~(Precision
IceCube Next-Generation Upgrade), ORCA~(Oscillation Research with
Cosmics in the Abyss) and INO~(India Based Neutrino Observatory) 
are utilizing resonant oscillation due to matter effect of atmospheric
neutrinos to achieve the same goal~\cite{IceCube:2013aaa,Ribordy:2013xea}.
As pointed out in Ref.~\cite{Blennow:2013vta}, due to the unprecedented
challenges in both types of experiments, each type alone might not
be able to reach the discovery sensitivity. However, as the two types
of experiments are constraining the key oscillation parameter $|\Delta
m_{32}^{2}|$ for MH resolution from different perspectives, they are
complementary thus combined data can increase the sensitivity to
MH. Fig.~\ref{fig:Combined-JUNO} taken from Ref.~\cite{Blennow:2013vta}
shows the $\Delta\chi^{2}$ for PINGU, JUNO (Daya Bay II) as well as combined
PINGU+JUNO+T2K  as a function of $\Delta m_{31}^{2}$\footnote{The choice 
of $|\Delta m_{31}^{2}|$ or $|\Delta m_{32}^{2}|$ does
not change the MH discussion as $|\Delta m_{31}^{2}|\gg\Delta m_{21}^{2}$.
It is only a change of the reference mass eigenstate.
} with the wrong MH assumed.

\section{\label{sec:Performance-requirements}Performance requirements of
MBRO LS detectors and R\&D}

\subsection{\label{sec:challenges}Summary of the unprecedented challenges in
MBRO experiments}

Based on the existing studies and our studies presented in previous
sections, we can summarize the major challenges of MBRO experiments as
follows,
\begin{itemize}
\item Statistics challenge. \\With 40~GW${_{\textrm{th}}}$ reactor power
and a detector of 20~kt LS at a baseline of 60~km, in 5 year running
time, a total of $\sim$$10^{5}$ inverse beta decay events can be
collected. Monte Carlo simulation shows that the $\chi^{2}$ difference
between the two MH hypotheses ($\Delta\chi^{2}$) fits for an Asimov
data set can reach $\sim$25 assuming a energy resolution of
$\sim$2.6\%/$\sqrt{E(MeV)}$ and perfect
systematics~\cite{Qian:2012xh}. However, based on MC,
the average probability to determine the correct MH is determined
to be 98.9\%~\cite{Qian:2012xh,Qian:2012zn}, not to  the conventionally
believed 5$\sigma$ level due to the different statistics followed by MH.
Thus in MBRO experiments, to resolve MH with high CLs, large statistics
is needed, which can only be achieved with massive detectors
together with powerful reactors. One of the major challenges in ensuring
sufficient statistics lies in the construction of a $>$10kt LS 
detector, which is unprecedented.
\item Reactor core distribution and site selection. \\MBRO experiments need
  multiple reactors to increase the statistics. However, if the multiple
  baselines differ by 1$\sim$5km, the MH sensitivity can be greatly reduced
  due to the cancellation effect~\cite{Yifang:2013,Li:2013zyd}.
  Also considering the backgrounds produced by the comic muons, the site needs
  to have an overburden $>\sim$500m (rock). JUNO has identified the Jiangmen
  site, which is $\sim$60km away from Yangjiang and the Taishan reactor
  complexes, to meet those criteria~\cite{Yifang:2013}. Suitable site locations
  are essential for the success of MBRO experiments.
\item Energy resolution. \\This is the well-recognized unprecedented challenge
in large LS detectors. We have showed that a $<\sim 3\% / \sqrt{E(MeV)}$
energy resolution is needed for the experiment, otherwise
the fine structure of the oscillation pattern due to MH is smeared
out, especially at low energy $<\sim$4 MeV. To achieve such a good
energy resolution, it requires dedicated R\&D programs to address the
following items ~\cite{Yifang:2012,Yifang:2013}: 
\begin{itemize}
\item High photo-cathode coverage of the detector, $\sim$80\% 
\item Photomultiplier tubes (PMT) with high collection efficiency and
  quantum efficiency (QE), $\sim$35\% 
\item Highly transparent LS with attenuation length of $\sim$35m (for the
  dimension of a 20kt LS detector)
\item High light yield LS ($\sim$1.5$\times$ photon yield of KamLAND LS)
\end{itemize}
\item Energy non-linearity. \\As shown in Ref.~\cite{Qian:2012xh},
  an unrecognized detector energy non-linearity could fake
  one MH as the other. Our sensitivity study in
  Sec.~\ref{sec:Medium-baseline-reactor-neutrino} shows that a degeneracy
  caused energy scale bias could greatly reduces the sensitivity. Therefore, to
  ensure the MH's discovery potential, the non-linearity of energy scale
  ($E_{rec}/E_{real}$) needs to be understood to a fraction of 1\% in a wide
  range of energy spectrum. This requirement should be comparable to  the
  current state-of-art 1.9\% energy scale uncertainty from
  KamLAND~\cite{2009KamLANDcal}. The non-linearity could come from different
  origins such as scintillation quenching, Cherenkov light, electronics,
  reconstruction non-uniformity, etc. An extensive calibration
  program is crucial for the experiment. Such a calibration program requires
  multiple energy calibration points inside the full fiducial volume of the
  detector. Multiple  types  of sources are desired, in particular positron
  sources to cover the wide range of the IBD spectrum to eliminate the
  potential degeneracy caused non-linearity. A dedicated R\&D program is
  definitely needed.
\end{itemize}

\subsection{\label{sec:pmt}PMT system study}
There are two major challenges for the PMT system in MBRO experiments. One
challenge of the PMT system is to ensure the mechanical safety and the other
is related to performance: low dark noise, high collection and quantum
efficiencies, and high photocathode coverage are needed. We consider the
following R\&D items are necessary: 
\begin{itemize}
\item PMT safety. \\Large format semi-hemispherical PMTs are leading
  candidates for the detector. Following the Super-K incident, the
  ability of PMTs to withstand hydrostatic pressure and control the
  propagation of PMT implosion impact has become critical
  issues for any large detector readout by
  PMTs~\cite{Diwan:2012zz,Diwan:2013zz}. PMT glass in liquid is known
  to undergo stress-induced corrosion. Even with good quality control,
  microscopic cracks in the PMT glass act as concentrators of stress. Once
  the concentrated stress exceeds the strength of glass, it will lead to
  cracking and failure (implosion). High photocathode coverage means
  high density of PMTs in the detector.
  In case of PMT implosion, the propagation of the shock waves to  the
  surrounding PMTs could cause chain action. 
  Therefore, a systematic study to contain the implosion impact  is
  needed to prevent the disaster.
  Based on our prior experience, a number
  of simulation studies must be performed to address PMT mechanical
  performance.
\item Collection and quantum efficiency enhancement. \\It is critical for the
  LS detector to reach $\ge$80\% photocathode coverage to control
  the energy resolution better than $\sim 3\% / \sqrt{E(MeV)}$. Due to
  geometric constraints, the design is quite challenging. Thus other means to
  enhance the detection efficiency are necessary. We have considered
  the following two options for this purpose:
  \begin{itemize}
  \item One option we have considered is to add Winston cones should the
      conventional PMTs are used by MBRO experiments. Winston cones
      have been utilized to increase light detection efficiency using specular
      reflection from metallic, mirror-like ellipsoidal surfaces and funneling
      light toward PMT photocathode surfaces. This is an attractive option to
      collect light lost in the spaces between PMTs. Mechanical designs need to
      be developed in order to accommodate Winston cones together with PMT
      mounting schemes.
    \item PMTs are susceptible to local magnetic field,
      such as the Earth's field of 0.4-0.5 Gauss. The magnetic field can reduce the PMT collection 
      efficiency as well as causing nonuniform response of PMTs to signals~\cite{DeVore:2013xma}.  
      Therefore, we need to R\&D various PMT magnetic field shielding methods to create  ideal
      working conditions for PMTs.
  \end{itemize}
\item PMT performance characterization. A new kind of PMT is being designed
  for JUNO using micro-channel plates~(MCPs) for PE collection
  and amplification. The performance of new PMT needs to be fully studied. 
\end{itemize}

\subsection{\label{sec:ls}Liquid scintillator study}

All studies from other groups and our group have shown it is absolutely
critical to reach energy resolution to less than 3\% at 1 MeV. 
One of the key elements in reaching the unprecedented energy
resolution in such a LS detector is higher photon yield. The typical
proposed dimensions are $\sim$35m in diameter for 
JUNO and $\sim$25m in diameter and height for RENO-50. Thus LS transparency
needs to reach $\sim$30m level at least. As we need large statistics to
resolve the MH, the aging rate of the LS to be sufficiently low so
transparency can stay at the acceptable level for at least 5 years.

The R\&D tasks of necessity are therefore:
\begin{itemize}
\item Purification of scintillator and fluor. We need to control scintillator
  impurities (e.g. non-radioactive chemical species that adversely affect
  optical properties) and develop methods to remove and assay residual
  radioactive contaminants, mainly from the naturally occurring $^{238}$U and
  $^{232}$Th decay chains. Typical liquid scintillator purchased
  from industry has an attenuation length of $\sim$10m with most impurities
  introduced during the production process. The best optical attenuation
  length achieved for a liquid scintillator is $L_{attn}\sim$20m at 430nm
  (LAB, after extensive purifications). Several technologies, such as vacuum
  distillation (SNO+, KamLAND, Borexino) or column extraction (industrial,
  petrochemical) can improve optical transmission. A combination of
  extraction column during the distillation phase might further improve the
  solvent. Another approach is the use of high-purity starting materials
  (e.g. nD instead of MO for LAB) as the feedstock for scintillator
  production such that cleaner liquid can be obtained. Other means of
  purification by separation resin or solvent washing should be tested. For
  instance, PPO (known to be a dirty material) can be cleaned by
  water-washing, recrystallization or solvent-distillation.

\item Search for a new scintillator that can be mass
  produced. Pseudocumene was selected by early experiments (Palo Verde, CHOOZ)
  due to its large scintillation output; however, it has a low flash point and
  poor material compatibility, thus a second, non-aromatic solvent (mineral
  oil or n-dodecane) has to be added to offset those effects. Such mixtures
  degrade the light-yield and complicate the scintillator handling (binary
  system). Instead, LAB with mild reactivity and high flash point (singular
  system) has been chosen by current generation experiments (Daya Bay, RENO and
  SNO+). LAB is the end result of extensive R\&D by SNO+, including a search of
  commercially available scintillators. A similar, improved survey
  of commercially available liquid scintillators is suggested.
  
\item Large Stokes-shift fluorophores. Most liquid scintillators have long
  optical transmission in the region of 440-550nm (e.g. LAB has
  $L_{attn}\sim$30m measured at 450nm emission) where the PMT still
  has good QE. Identification of fluor/shifter combination optimizing
  in higher optical emission and light-yield by comparing their fluorescence
  emission and intrinsic light-yields in different scintillators are the
  main objectives for this task.
  
\item Temperature effects and time dependent characteristics. Significant
  improvement in the liquid scintillator light yield could be achieved by
  lowering the LS temperature (e.g. $\sim$5\% more light at 10$\celsius$
  compared to 20$\celsius$ measured by IHEP). Confirming the
  temperature dependence and studying time characteristics of LS
  along with the potential for pulse shape discrimination of different
  scintillator samples are needed. Pulse shape discrimination may play an
  important role in intrinsic background rejection in large liquid scintillator
  detectors. We can finalize the LS temperature and pressure dependence
  effects utilizing previous experience and upgraded apparatus previously
  used for Hanohano R\&D. Such measurements can be one of the key inputs
  for detector design and operation. 
  
\item Light-yield Measurements. It is well known that quenching affects light
  output and can lead to reduced scintillation efficiency. The light-yield of
  surveyed scintillators can be screened by different radioactive
  sources: $\alpha$, $\beta$, and $\gamma$. The selected scintillators'
  light yield as a function of electron energy can
  then be studied systematically. The energy response of the liquid
  scintillator is then to be used for MC simulations.
  
\item This detector is likely to be the largest scintillator detector built in
  the next 5 years. One potential problem is that laboratory equipment is
  normally not comparable to the actual size of detector. This creates
  ambiguity when extrapolating laboratory measurements to the real detector;
  for instance, both Daya Bay and KamLAND observed 10$\sim$15\% more
  scintillation light in the detector than small lab modules. This is likely
  due to absorption/re-emission or scattering of light propagating through
  the scintillator. Thus, a satisfactory R\&D program needs to build a
  one-dimensional tube with length 
  close to the detector size to (1) investigate light propagation mechanisms
  as a function of path-length and (2) verify PE yield at
  $\sim$30m. 

\end{itemize}

\subsection{\label{sec:elec}Front-end and trigger electronics}
The requirements and specifications of the front-end electronics and trigger
system, in particular, the linearity of the charge measurement, dynamic
range of time and charge measurements, multiple hit and pileup resolution
capability, waveform digitization frequency, types of triggers and their
implementation are critical parts of the experiment.

The MBRO LS detector will be a large device with approximately
20,000 PMTs that need to be read out. The readout should address the issue
of accurate measurement of low level signals of interest while
simultaneously preserving information of large signals originating from
muons interacting in the detector. As most of the
proposed front end readout will be located inside the detector, reliability
of connections and electronics is critical. The development of a readout
chain that can be immersed in the detector, digitize and group signals close
to the photodetectors and transport the digital information via optical fibers
to the DAQ would be desired.

The U.S. team has successfully developed electronic systems meeting such
needs in the past. A dynamic range compressor that can preserve
the linearity of both low and high amplitude signals is shown in
Ref.~\cite{cleland:1996}. The design will preserve the
number of channels and reduce the number of cables that will carry signals
out of the detector for further processing. 
The digitizer for the fifth generation TeV Array with Gsa/s sampling and
Experimental Trigger (TARGET)~\cite{Bechtol:2011tr} can be used in
MBRO experiments. It is an ASIC that contains 16 channels of transient
waveform recorder that was designed to be used in highly pixelated photon
detectors for large neutrino and muon detectors. 

The details of how to design and deploy a reliable system is central to such
R\&D efforts. In the past, the U.S. team has developed electronics deployed
in other experiments with little to no access and we are quite confident
that a solution to achieve the necessary reliability is possible. 

\subsection{\label{sec:calibration}Detector energy response calibration}

As one of the most crucial requirements, detector energy response
calibration should be one of the U.S. R\&D program's main prioritized
focuses. We have considered various options in calibrating the detector using
a combined approach based on our experiences obtained in LS detector
experiments like KamLAND and Daya
Bay\ \cite{2009KamLANDcal,2013arXiv1305.2248L,2013arXiv1305.2343H}. Due to
the size of the MBRO detectors, uniformity plays a more significant
role than smaller detectors which have been constructed in the past. We
are considering options of dissolving short-lived radioactive
isotopes to calibrate the detector response besides the
mature approaches using the uniformly distribution intrinsic events
like spallation neutrons. To have precise position dependence studies,
we are studying multiple ways of deploy point calibration sources
to the whole volume. We know that different types
of particles exhibit different energy responses in a LS detector, thus it is
critical to select the most suitable calibration sources to reduce the
non-linearity uncertainties. We are also studying different types of
radioactive sources.

We learned from Daya Bay and KamLAND experiments that multiple energy scale
models can generally be constructed with different types calibration samples
and different emphases. The various models on one hand explore different
aspects of the detector, which is good for us to fully understand our
detector. On the other hand, at the same time, it unavoidably increases the
energy scale uncertainties and functional form complexities which decrease
the sensitivity of the experiment to MH as shown in
Sec.\ref{sub:Sensitivity-to-mh}. The complementarities or conflicts
depending on our points of view, among various models are largely due to the
lack of positron calibration sources and the lack of continuous coverage
over the entire IBD spectrum. In view of this, we are considering to build a
positron accelerator on-site which would provide  mono-energetic positrons to
calibrate the detector for IBD events. After reviewing
different technologies, we think  positron pelletron is reliable and would be
able to deliver mono-energetic positrons that cover the IBD spectrum
continuously. Its dimension is on the order of $\sim$10m thus can fit in JUNO
or RENO-50 type experimental halls easily. Based on survey of existing
facilities, 0.5 to 6.5 MeV positron beams can be produced with an accuracy
of $\Delta E/E \leq 10^{-4}$~\cite{Bauer:1990zz}. The pelletron can also
switch to use electron sources thus provide electron beams in the same
energy range.

With both the calibration source and the tunable positron/electron pelletron
approaches, we believe a highly accurate energy response model of MBRO LS
detectors can be obtained to meet the needs of MH sensitivity and precision
measurements. Naturally, the following R\&Ds are needed in order to mount a
successful calibration system for MBRO experiments:

\begin{itemize}
\item Develop a liquid scintillator test chamber for the precise laboratory
  study of the light response and non-linearity of the liquid scintillator
  with the capability to characterize in-situ calibration
  sources. This will provide a precise characterization of calibration
  sources in the LS and allow tests of secondary effects from shielding and
  source encapsulation.
\item Develop concepts for the deployment and precise positioning of
  radioactive and light sources throughout a 30m large detector. 
\item Develop a concept for the injecting and distributing of uniformly
  distributed short-lived
  radioactive isotopes for calibration of the detector volume.
\item Develop a tunable positron/electron gun in the energy range of
  1-8 MeV for continuous calibration over the reactor antineutrino
  spectrum. Pelletron technology is an attractive option for this purpose.
\end{itemize}

\section{\label{sec:Other-potential-physics}Other potential physics topics}

With 10-20~kt LS detectors underground, MBRO experiments offer many
other physics opportunities. We only list a few highlights below:
\begin{itemize}
\item Supernova Neutrinos:\\ It is estimated that for a typical
  ($3\times10^{53}$ erg) supernova 10~kpc away, $\sim$3000 inverse beta
  decay events can be detected, together with $\sim$3000 events in other
  visible channels.
\item Geoneutrinos:\\ With a 20~kt LS detector, JUNO expects to observe
  $\sim$750 geoneutrino events per year. The reactor background in the
  energy range of geoneutrinos ($1.8-3.3$ MeV) is about 4 times larger. The
  difference in shape and time dependence will help to extract the
  geoneutrino signals.
\item Proton Decay:\\ If nanosecond timing resolution can be reached, MBRO
  LS detectors will be in good position to measure proton decay
  in the decay channel $p\to K^{+}\bar{\nu}$. Scaling from LENA's
  estimation~\cite{Wurm:2011zn}, a lower limit of proton lifetime,
  concerning the decay channel investigated, of $\tau>2.4\times10^{34}$ y
  (at 90\% C.L.) could be reached in 10 years by JUNO.
\end{itemize}

\section{Summary and conclusions}

Medium-baseline reactor neutrino experiments using massive LS detectors
provide a unique opportunity to determine the neutrino mass hierarchy. Such
experiments have unprecedented detector performance requirements which have 
not been realized before. Our independent calculation confirms the 
sensitivities claimed by other groups in the form of $\Delta\chi^{2}$
values under near ideal condition with different assumptions on the LS
detector performance. In addition to the widely recognized and
accepted no-go condition of energy resolution worse than $\sim 3\% /
\sqrt{E(MeV)}$, we find that understanding the uncertainty of MBRO
LS detectors' energy non-linearity plays a crucial role in its
sensitivity to MH. Under certain types of non-linear bias, sensitivity will
be significantly reduced. A dedicated energy scale calibration system must
be developed to control the energy scale uncertainty.

We find that a good knowledge of the reactor spectrum can relax the 
requirements on the energy calibration as the spectrum is correlated
between different energies. If understandings on reactor spectrum
won't be improved significantly during the course of MBRO experiments,
Fourier method or the spectrum ratio method are less sensitive to the
reactor spectrum but 
the requirement in the absolute energy scale becomes more
stringent. Under such scenarios, we recognize that a second detector
at a suitable baseline $\sim$30km can mitigate the stringent
requirement on energy scale thus improve the sensitivity.

Besides experimental challenges, we also find
due to the discrete nature of MH, the $\sqrt{\Delta\chi^{2}}$ rule
of setting measurement confidence levels does not apply in MH determination.
Combining all factors, we find the best scenario of current MBRO experiment
proposals is to reach $2\sim 2.5 \sigma$ sensitivity in MH ($\Delta
\chi^2 =16\sim25$).

MBRO experiments offer rich physics programs which extend from precision
neutrino oscillation parameter measurements, planetary science to the
observation of geoneutrinos, test of grand unification theories via proton
decay studies and, potentially, study astrophysics via ex-territorial
neutrinos. The oscillation parameters' precise measurements are guaranteed
to reach sub-percent level should we reach the designed performance of MBRO
detectors. Such precision results would enable a future direct
unitarity test of the neutrino mixing PMNS matrix's to $\sim$1\%
level. Recent studies show that combining the data from MBRO experiments
with next generation long-baseline beam neutrino and atmospheric neutrino
experiments like NO$\nu$A, T2K, INO, PINGU, and ORCA can enhance the
sensitivity to MH. The power lies in that the atmospheric mass-squared
splitting is constrained systematically different in MBRO experiments and in
beam/atmospheric neutrino experiments. The forthcoming neutrinoless double
beta decay experiments will also benefit from precise measurements of the
solar mixing angle as it helps to improve the constraints on absolute
neutrino mass $\langle m_{\beta \beta} \rangle$. In order to reach the
desired accuracy in the MBRO LS detectors' energy responses, we
have considered various calibration strategies. By combining radioactive
sources deployed in multiple ways and a positron pelletron with tunable
mono-energetic positron beams, we can calibrate the detector with complete
coverage in both spatial aspect and energy spectrum aspect.

MBRO experiments face many unprecedented challenges while offering great
physics opportunities. Well-designed R\&D programs are needed to ensure its
success. The U.S. team is experienced in many respects of such experiments
thus is in a very good position to initiate a US R\&D program for MBRO
experiments. The current estimated cost of JUNO in China is a few hundred
million dollars with an estimated 5 years for construction. Primary support
for this experiment from Chinese funding agencies looks very promising and
there are substantial opportunities for international collaboration. R\&D
and site investigations in China are underway. Data taking could begin
around 2020. A similar proposal named RENO-50 has been proposed in South
Korea with slightly shorter baseline and 90\% of the JUNO's target mass. A
U.S. R\&D program will ensure the U.S. team master the key technologies of
the critical experimental components of MBRO experiments and make
irreplaceable contributions in the forthcoming MH discovery and precision
neutrino physics era.

\bibliographystyle{jhep}
\bibliography{ref}

\end{document}